\documentclass[aps,prl,twocolumn,superscriptaddress,showpacs]{revtex4}
\usepackage{graphicx,amsmath,amssymb}
\usepackage{color}
\usepackage[normalem]{ulem}	

\begin{document}

\title{The magnetic precursor of the pressure-induced superconductivity in Fe-ladder compound}
\author{Songxue Chi}
\affiliation{
Quantum Condensed Matter Division, Oak Ridge National Laboratory, Oak Ridge,
Tennessee 37831, USA
}
\author{Yoshiya Uwatoko}
\affiliation{
Institute for Solid State Physics (ISSP), University of Tokyo, Kashiwa, Chiba
277-8581, Japan }

\author{Huibo Cao}
\affiliation{
Quantum Condensed Matter Division, Oak Ridge National Laboratory, Oak Ridge,
Tennessee 37831, USA
}
\author{Yasuyuki Hirata}
\affiliation{Institute for Solid State Physics, The University of Tokyo, Kashiwanoha 5-1-5, Kashiwa, Chiba 277-8581, Japan}
\author{Kazuki Hashizume}
\author{Takuya Aoyama}
\affiliation{Department of Physics, Graduate School of Science, Tohoku University, 6-3, Aramaki Aza-Aoba, Aoba-ku, Sendai, Miyagi 980-8578, Japan}
\author{Kenya Ohgushi}
\affiliation{Department of Physics, Graduate School of Science, Tohoku University, 6-3, Aramaki Aza-Aoba, Aoba-ku, Sendai, Miyagi 980-8578, Japan}
\affiliation{Institute for Solid State Physics, The University of Tokyo, Kashiwanoha 5-1-5, Kashiwa, Chiba 277-8581, Japan}

\begin{abstract}
The pressure effects on the antiferromagentic orders in iron-based ladder compounds CsFe$_2$Se$_3$ and BaFe$_2$S$_3$ 
have been studied using neutron diffraction. 
With identical crystal structure and similar magnetic structures, 
the two compounds exhibit highly contrasting magnetic behaviors under moderate external pressures.    
In CsFe$_2$Se$_3$ the ladders are brought much closer to each other by pressure, but the stripe-type 
 magnetic order  shows no observable change.
In contrast,  the stripe order in BaFe$_2$S$_3$, undergoes a quantum phase transition where 
an abrupt increase of N$\acute{e}$el temperature by more than 50$\%$ occurs at about 1 GPa, accompanied by a jump in the ordered moment. 
With its spin  structure unchanged, BaFe$_2$S$_3$ enters an enhanced magnetic phase that bears the characteristics 
of an orbital selective Mott phase, which is the true neighbor of
superconductivity emerging at higher pressures.

\end{abstract}


\maketitle

The antiferromagnetic (AF) phase adjacent to superconductivity (SC) is so richly faceted
that its microscopic origin still eludes a unified description. Significant variation of the ordered magnetic moment and the 
underlying degree of electron correlations lie at the heart of the heated dispute \cite{Dai,Si,Yin}.
The static AF phase in the parent compounds has roughly two categories: stripe magnetism and block magnetism. The former includes the single stripe 
in LaFeAsO, BaFe$_2$As$_2$, NaFeAs, and double stripe in FeTe \cite{Pdai15}. Spin block 
order was found in the vacancy-ordered K$_2$Fe$_4$Se$_5$ (245) \cite{Bao11}. These materials all have a plane of Fe square lattice once deemed 
indispensable for the occurence of SC. 
The recent successful induction of SC by pressure in the ladder compound 
BaFe$_2$S$_3$ \cite{Takahashi,Yamauchi} has introduced a quasi-one dimensional structural motif for the studies of 
iron-based superconductors and a parallel to the quasi-1D cuprates \cite{Uehara}.
As if the layers of the superconducting Fe square lattice were sliced up and staggered, 
the \textit{A}Fe$_2$\textit{X}$_3$ (\textit{A} = K, Rb, Cs or Ba and \textit{X} = Chalcogens) compounds consist of ladders of two-leg 
Fe-chains with edge-sharing tetrahedra of anions (Se or S) 
surrounding each Fe site, as shown in Figure 1(a). 
The reduced dimensionality leads to modified bandwidth \cite{Caron12}, Fermi surface topology, 
and provides a rare insight into critical open issues such as the nature of the AF order.

Both stripe- and block-types of AF
orders are hosted by the Fe-ladder compounds. BaFe$_2$S$_3$ and
CsFe$_2$Se$_3$, with the CsCu$_2$Cl$_3$ type structure ($Cmcm$ space group), have the stripe AF order
where the ferromagnetic (FM) spin pairs on the same ladder rung correlate antiferromagnetically 
along the leg direction. The ordered 
moment lies in the rung-direction in BaFe$_2$S$_3$ (Fig.1(b))\cite{Takahashi} and the leg-direction in CsFe$_2$Se$_3$ (Fig.1(c)) \cite{Du12}.
In BaFe$_2$Se$_3$, the distorted FeSe$_4$ tetrahedron lose the $C$-centering and result in the lowered symmetry $Pnma$ \cite{Hong}.
The magnetic structure consists of blocks of 4 FM spins forming alternating 
AF pattern along the leg direction \cite {Nambu}. The magnetic excitations in BaFe$_2$Se$_3$ 
fits the description of localized spins and an orbital-selective Mott phase \cite{Mourigal}.

\begin{figure}[b!]
\label{fig1}
\includegraphics [width=1.0\columnwidth]{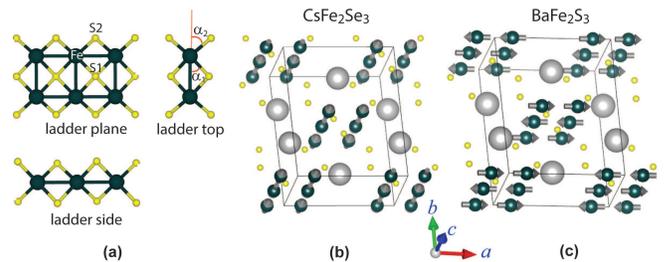}
\caption{(color online) (a) The structure of the Fe-ladder and its relative positions with anions (Se or S) for 
the ladder compounds adopting the $Cmcm$ space group.  
The magnetic structure in (b) CsFe$_2$Se$_3$  with spins parallel to 
the $c$-axis and (c) in BaFe$_2$S$_3$ with ordered moment along $a$. 
}
\end{figure}

The pressure-induced metal-insulator transition in BaFe$_2$S$_3$ is categorized as a bandwidth-control type Mott transition \cite{Takahashi,Yamauchi}. 
The AF order is suppressed before SC arises at higher pressures \cite{Yamauchi}.  
To elucidate the SC pairing mechanism, the detailed evolution of the  AF phase under pressure is the crucial step still missing.
In this Letter, we present a pressure effect study on the AF orders in 
the single crystalline  CsFe$_2$Se$_3$ and BaFe$_2$S$_3$ using
neutron diffraction.  
The two compounds contrast in ladder spacings and electronic properties. We show that they also  
exhibit highly contrasting responses to pressures.    
The magnetism in CsFe$_2$Se$_3$ is  robust against the applied pressures close to 2 GPa. 
The AF order in BaFe$_2$S$_3$ undergoes a rather abrupt enhancement around 1 GPa, both in transition temperature and ordered moment, before being suppressed at higher pressures. Such unusual change qualifies as an orbital selective Mott transition.

Single crystals of BaFe$_2$S$_3$ and CsFe$_2$Se$_3$  were prepared by the 
solid-state reaction method \cite{Takahashi}.
The samples were inserted into Teflon capsules and loaded in a piston cylinder cell made of CuBe alloy or Zr-based metallic glass \cite{Komatsu}.
Daphne oil was used as the pressure transmitting medium. 
The single crystal neutron diffraction measurements
were carried out on HB-3A Four-circle
Diffractometer at the High Flux Isotope Reactor (HFIR) of the
Oak Ridge National Laboratory (ORNL). The wavelengths of 1.003 \AA{}  and 1.542 \AA{} were
employed.   
The pressures were calibrated with NaCl single crystal loaded together  with the sample in the cell \cite{SM}. 
One of the applied pressures was calibrated on the Wide Angle Neutron Diffractometer (WAND) at HFIR. The Rietveld
refinements on the crystal and magnetic structures
were conducted using the FullProf Suite \cite{Fullprof}. 

\begin{figure}[b!]
\label{fig2}
\includegraphics [width=1.0\columnwidth]{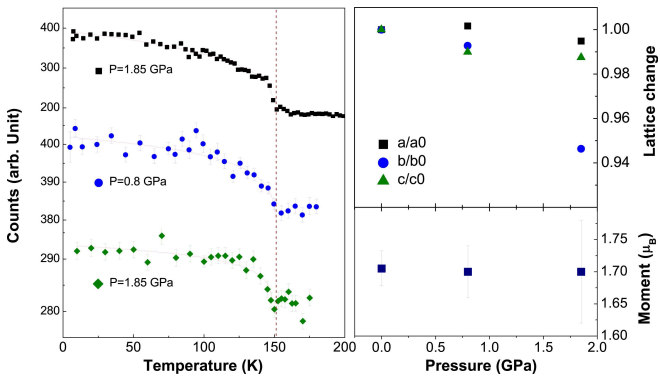}
\caption{(color online) The effect of pressure on structural and magnetic properties in CsFe$_2$Se$_3$. (a) The temperature dependence of the (0.5,2.5,1) peak intensity at ambient pressure, 0.8 GPa and 1.85 GPa. 
(b) The change of lattice constants under various pressures. (c) The size of the ordered moment at different pressures.
 }
\end{figure}
 
We report the structural information at 4 K. Both compounds can be well-described by the orthorhombic space group $Cmcm$. The lattice constants of CsFe$_2$Se$_3$ are $a$=9.7105(9) \AA{}, $b$=11.595(1) \AA{}, and $c$=5.6659(3) \AA{}. The lattice constants and structrual parameters of BaFe$_2$S$_3$ are summarized  in Table I. The biggest contrast is in $a$, which means the ladders in the same row are closer in BaFe$_2$S$_3$ since the two compounds have almost the same rung length. 
The ladder leg is bigger in CsFe$_2$Se$_3$. Moderate hydraulic pressure changes the spacings between the ladders and does little to the size of the ladders. 

\begin{table}[ht!]
\caption{Structural parameters for BaFe$_2$S$_3$ at 4 K at ambient pressure (top) and 1.3 GPa (bottom). The ambient pressure lattice constants are $a$=8.8607(4) \AA{}, $b$=11.2767(6) \AA{}, and $c$=5.2730(6) \AA{}.  The length of the ladder rung is 2.727 \AA{} and that between rungs is $c$/2=2.6365 \AA{}. Those for 1.3 GPa are $a$=8.6172(4) \AA{}, $b$=11.0169(1) \AA{}, and $c$=5.2159(5) \AA{}.  The length of the ladder rung is 2.701 \AA{} and that between rungs is $c$/2=2.608 \AA{}.
}
\label{tab1}
\begin{ruledtabular}
\begin{tabular}{llccccc}
Atom & site & \textit{x}  & \textit{y}  & \textit{z}  \\
		 \hline
Ba & 4\textit{c} & 1/2 & 0.181(4) &1/4   \\
Fe & 8\textit{e} & 0.3461(8) & 1/2 & 0   \\
S(1) & 4\textit{c} & 1/2 & 0.612(9) &1/4  \\
S(2) & 8\textit{g} & 0.2091(2)& 0.364(5) & 1/4   \\
		 \hline
Ba & 4\textit{c} & 1/2 & 0.185(3) &1/4  \\
Fe & 8\textit{e} & 0.343(2) & 1/2 & 0   \\
S(1) & 4\textit{c} & 1/2 & 0.613(7) &1/4  \\
S(2) & 8\textit{g} & 0.210(7)& 0.374(5) & 1/4    \\
  \end{tabular}
\end{ruledtabular}
\end{table}

At ambient pressure, the magnetic reflections for both compounds were collected using the propagation wave vector (1/2,1/2,0). Representation analysis provides four different irreducible representations (irreps) $\Gamma_1$, $\Gamma_2$, $\Gamma_3$ and $\Gamma_4$, each of which consists of 3 basis vectors (BV)\cite{SM}.  We sort through all BVs in each irrep for refinement and obtained the best $R$-factor from $\phi_9$ for CsFe$_2$Se$_3$ and $\phi_1$ for BaFe$_2$S$_3$. The ordered moment of  1.705(27) $\mu_B$ lies along the $c$-direction in CsFe$_2$Se$_3$ [Fig. 1(c)]. The magnetic peak intensity
as a function of temperature was fitted to a power law, plotted as the red curves in Fig. 2(a), which gives the N$\acute{e}$el temperature, $T_N$.  $T_N$ is estimated to be 149 K. These findings are all consistent with the
previous powder diffraction study \cite{Du12}.
The refined moment in BaFe$_2$S$_3$ is 1.043(30) $\mu_B$ along the $a$-direction,  as shown in Fig. 1(c), which is smaller than the reported 1.20(6) $\mu_B$ in ref.\cite{Takahashi}. Its AF transition temperature, $T_N$=105 K [Fig. 3(a)], is also lower than the reported 119 K in 
Ref.\cite{Takahashi}. The slightly weaker AF order in the present sample can be explained by the strong dependence of magnetic properties on the synthetic procedure \cite{Takahashi}. 

For both compounds, identical crystals were pressurized for the pressure measurements. 
We first discuss the effect of pressure on CsFe$_2$Se$_3$, as summarized in Fig. 2. The magnetic wave vector remains unchanged up to the highest 
applied pressure (1.85 GPa). The magnetic intensity at (0.5,2.5,1) develops about the same temperature at 0.9 GPa (152 K) and 1.85 GPa (150 K) as the ambient pressure (149 K). 
Rietveld fits 
confirmed the unchanged nuclear structure and spin configuration under the two pressures. The size of the ordered moment also remains the same [Fig. 2(c)].
The lattice constants decrease at different rates under pressure. At 1.85 GPa, 
$a$ and $c$ decrease by less than 2$\%$, but $b$ decreases by more than 5$\%$ and becomes 11.22 \AA. The distance between 
the ladder stacking layers in CsFe$_2$Se$_3$ under 2 GPa is even slightly smaller than that in unpressurized BaFe$_2$S$_3$.

\begin{figure}[bt!]
\label{fig3}
\includegraphics[width=1.00\columnwidth]{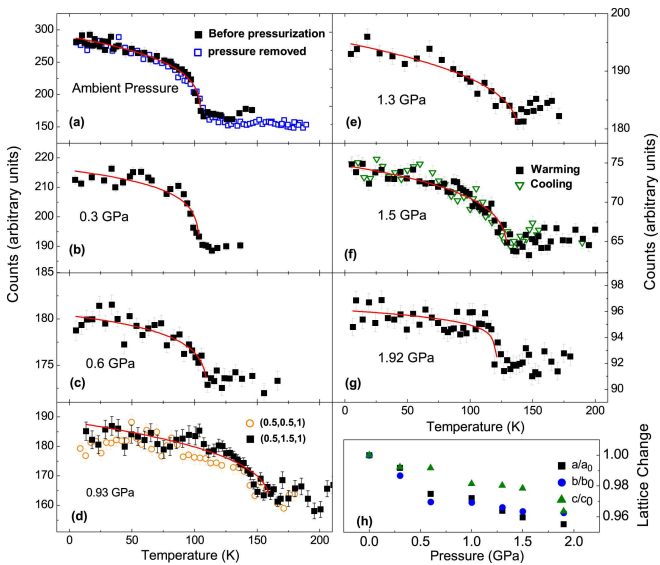}
\caption{(color online) The temperature variation of the magnetic peak intensities in BaFe$_2$S$_3$ at different pressures. (a) (0.5,1.5,1) at the ambient pressure before the hydraulic pressure is applied and after the pressure has been removed (b) (0.5,2.5,1) at the pressure of 0.3 GPa (c) (0.5 1.5,1) at 0.6 GPa
(d) (0.5,0.5,1) (orange open circle) and (0.5,1.5,1) (black solid square) at 0.93 GPa. The power law fit is for (0.5,1.5,1)  (e) (0.5,1.5,1) at 1.3 GPa (f) (0.5,1.5,1) on warming and cooling at 1.5 GPa and (g) (0.5,1.5,1) at 1.93 GPa. (h) The change of lattice parameters as a function of pressure. 
}
\end{figure}

In contrast to the strong magnetic order in CsFe$_2$Se$_3$, the magnetic phase in BaFe$_2$S$_3$ exhibits 
remarkable sensitivity to pressures. Fig. 3(b) shows the order parameter at the pressure of 0.3 GPa. $T_N$ is estimated to be 104 K, which implies that
the AF order is unaffected. At 0.6 GPa, (0.5,0.5,0) remains as the magnetic propagation wavevector. $T_N$ shows a slight 
increase to 112 K [Fig.3(c)]. The Rietveld refinements using intensities of rocking curve scans collected at 0.6 GPa show no major
change of crystal and spin structures. The variation of refined moment, 1.02(8) $\mu_B$, from the ambient pressure 
value is smaller than the statistical error.

As pressure is increased to 0.95 GPa, a drastic change of the magnetic order occurs. 
The change of (0.5,1.5,1) intensity on warming shows that the magnetic transition becomes 164 K, a leap of 56$\%$ from the ambient pressure 
and 47$\%$ from 0.6 GPa [Fig. 3(d)]. The increase of $T_N$ at such a rapid rate, 132.5 K/GPa, is unprecedented. 
To confirm this dramatic effect of pressure we perform the same temperature measurement on another magnetic reflection (0.5,0.5,1), as represented by the orange open circle in Fig. 3(d), which shows the same $T_N$.
To obtain the correct spin structure of the pressure-strengthened magnetic phase, 
we made broad surveys on the potential magnetic reflection positions. Using 
the area detector and varied temperature we ruled out the  
possibility of a different magnetic wave vector \cite{SM}. 
All the real magnetic peaks were found at the ($m/2,n/2,l=1$) ($m$ and $n$ are intergers) positions. 
The Fullprof refinement using these peaks yielded the best $R$ factor with the same $\phi_1$ of $\Gamma_1$, an unchanged structure,  and revealed that the ordered moment jumped to 1.24(5) $\mu_B$. 

Further increase of pressure immediately starts to suppress $T_N$.  
It decreases to 139 K at 1.3 GPa [Fig. 3(e)] and to 131 K at 1.5 GPa [Fig. 3(f)]. The order parameter on cooling shows no hysteresis, suggesting the glassy behavior in Ba$_{1-x}$Cs$_x$Fe$_2$Se$_3$ \cite{Hawai} and Ba$_{1-x}$K$_x$Fe$_2$S$_3$ \cite { Hirata} is likely caused by the change of carrier concentrations.
Fig. 3(g) shows a continued suppression
of $T_N$ to 119 K at 1.93 GPa.
We carried out  refinements for all pressures above 0.95 GPa, which show that the same stripe type of magnetic order and the same value of 
moment persists to the highest measured pressure. After depressurization, the order parameter measurement was taken on the same sample which
shows the original value of $T_N$, as shown by the open blue square in Fig. 3(a).

To obtain more information on the crystal structure at pressures above the sharp change in AF order, 
we used a pressure cylinder made of Zr-based metallic glass for \textit{P} = 1.3 GPa. The material does not produce sharp Bragg reflections \cite{Komatsu} and allowed us to collect more Bragg reflections
from the sample. The refined structural parameters at 1.3 GPa, together with those at ambient pressure, are summarized in Table I.

\begin{figure}[bt!]
\label{fig4}
\includegraphics[width=1.0\columnwidth]{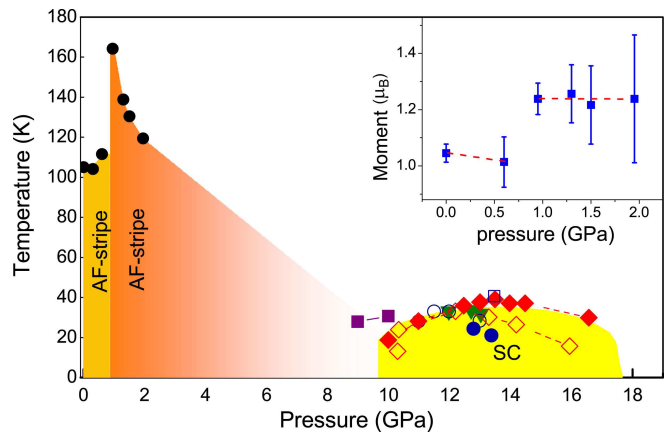}
\caption{(color online) Pressure-temperature phase diagram of BaFe$_2$S$_3$ showing the AF and superconducting transitions. Black solid circles correspond with the result of present study while other symbols represent the results from Ref. \cite{Takahashi}. The inset shows the size of ordered moment of
the stripe magnetic phase as a function of pressure.
 }
\end{figure}

Our neutron results of the magnetic evolution in BaFe$_2$S$_3$ under hydraulic pressure are summarized in the $P-T$
phase diagram in Fig. 4, along with the pressure-induced SC phase from Ref.\cite{Takahashi}.
The refined moment sizes at various pressures are shown in the inset of Fig. 4. For pressures higher than 2 GPa, we 
know the suppression of the magnetic order continues untill the 
SC starts\cite{Yamauchi}. We separate the AF phases below and above 0.95 GPa with two colors, the boundary of which represents
a pressure-induced magnetic phase transition manifested by a tremendously boosted $T_N$, accompanied by a jump in the ordered moment.
SC occurs in this $Cmcm$ ladder structure \cite{Takahashi}, thus this pressure-enhanced stripe-type AF order
is the true of precursor of the spin fluctuations that might correspond with the SC state. 

The absence of structural transitions under pressure in these two ladder compounds is to be expected because the $Cmcm$ phase is very stable. 
Both pressure and heating can drive a $Pnma$ to $Cmcm$ structural transition
in BaFe$_2$Se$_3$ \cite{Svitlyk}. Such a transition can also be achieved by chemical pressure as in Ba$_{1-x}$Cs$_x$Fe$_2$Se$_3$ \cite{Hawai}. 
The $Cmcm$ phase has two nonequivalent anion sites, in case of BaFe$_2$S$_3$, S1 and S2. As shown in Fig. 1(a),
S1 is between the ladder legs and S2 is out of the ladder. Not only do the two S sites have different distances from Fe, but also
different heights above the ladder plane.
This makes the point symmetry surrounding the Fe ions $C_s$ instead of $S_4$ as in 
the 2D Fe compounds. Such deviation implies different crystal field schemes and orbital states in the $Cmcm$ ladder compounds.
The differences between the two S sites are further increased by pressure in BaFe$_2$S$_3$. Compared to ambient pressure, the Fe-S1 distance at 1.3 GPa decreases from 2.285  
to 2.258 \AA{} and Fe-S2 decreases from 2.269 to 2.237 \AA{}. The angles $\alpha_1$ and $\alpha_2$, as defined in Fig. 1(a),
changes from 43.65$^{\circ}$ to 42.90$^{\circ}$ and from 48.55$^{\circ}$ to 49.55$^{\circ}$, respectively. The change in the size the Fe ladder 
is smaller than the standard error.

The valence of Fe ions in CsFe$_2$Se$_3$ is supposed to be a formal mixed +2.5. However Mossbauer \cite{Du12} and photoemmission \cite{Ootsuki} studies indicate that all the Fe sites take the Fe$^{2+}$ configuration and the Se $4p$ holes are trapped at the Se sites 
between the two legs \cite{Ootsuki}. The localized Se $4p$ holes and thus the Fe $3d$ electrons make CsFe$_2$Se$_3$
a charge-transfer-type Mott insulator, and are essential in stabilizing the stripe-$c$ magnetic phase.
Substituting Ba with K in BaFe$_2$Se$_3$ \cite{Caron12} results in the switch from the block magnetic phase to stripe-$c$ phase. Similarly, doping Cs 
\cite{Hawai} triggers a switch from block to stripe-$a$, then to stripe-$c$. In both cases, the transition to stripe-$c$ order is accompanied by
an increase in variable range hopping, indicating the localization of carriers.
The dominating transfer integral is between the nearest neighbor $d_{3z^2-r^2}$ orbitals \cite{Arita, Du12},  which is along the leg direction. 
The long Fe-Fe bond distances along the leg direction, u=2.83 \AA{} in CsFe$_2$Se$_3$, as opposed to u=2.63 \AA{} in BaFe$_2$S$_3$, also helps to localize the charges and stabilize the stripe-$c$ order. The pressure does not shorten the rung enough to
disturb the magnetic order even though the ladders are brought much closer. This indicates that the inter-ladder exchange interaction and transfer integral 
in the ladder compounds are small.

In comparison, BaFe$_2$S$_3$ is not an insulator but a semiconductor with a small energy gap of 0.06-0.07 $eV$ \cite{Reiff, Gonen}. Localized Fe $3d$ electrons coexist with  itinerant electrons \cite{Ootsuki}.
In a localized regime, if the exchange interaction is affected by 
the pressure through the compressed lattices, the pressure dependence of the AF transition temperature follows the Bloch's rule \cite{Bloch}.
The smooth variation of lattices would result in a gradual increase of $T_N$, as in Fe$_3$O$_4$ \cite{Samara} and La$_{1.4}$Sr$_{1.6}$Mn$_2$O$_7$ \cite{Kamenev}. This is certainly not the case for BaFe$_2$S$_3$. In an itinerant picture, on the other hand,
the pressure can potentially modify the Fermi pockets \cite{Arita} to improve the nesting feature, but neither hole- nor electron-doping
produces such drastic magnetic enhancement \cite{Reissner,Hirata}.
If the pressure indeed reduces correlation, by increasing the bandwidth and decreasing $U$ \cite{Arita}, and subsequently delocalizes Fe $3d$ electrons, the increased hopping on the ladder rungs, through some double exchange mechanism, should be able
to enhance the FM interactions. This would explain the increased $T_N$, but not the increased moment. 

The simultaneous spring of $T_N$ and the ordered moment at about 1 GPa
signals a quantum phase transition (QPT) that eludes first principle studies \cite{Suzuki,Arita, Luo13}.
This QPT ushers the BaFe$_2$S$_3$ system into the true Mott phase whose gap closes at higher pressures to pave the way for
the SC phase \cite{Yamauchi}.
It has the apparent fingerprints of an orbital selective Mott transition (OSMT): 
(i) The unchanged magnetic structure and spin orientation rule out the possibility of a metamagnetic transition.
(ii) The change of moment and its two-stage saturation has been predicted by the theories of OSMT \cite{deMedici09, Werner07, Werner08,Yin, Rincon14}.
In these theories, change of occupancies of $3d$ orbitals brings half-filled $t_{2g}$ shell that can be readily localized.  
In our case, the sulfur tetrahedron modified by pressure may increase the crystal field splitting,
which in turn changes the orbital occupancies. With a robust Hund's interaction that decouples bands \cite{deMedici11b}, 
the localization only needs to happen to one of five $3d$ orbitals, all of which contribute to the AF order \cite{Suzuki}. (iii) The
maximum value of $T_N$ at the pressure-induced QPT is also characteristic of Mott critical coupling
under the influence of strong Hund's rule coupling \cite{deMedici11,Mravlje}.  
(iv) An unknown transition at about 200 K  \cite{Hirata,Yamauchi}
tends to decrease and merge with $T_N$ as pressure increases \cite{Ootsuki,Yamauchi}. 
This transition is possibly related to orbital ordering and hints of
the critical role of the orbitals in forming the magnetic ground state in BaFe$_2$S$_3$.

In summary, moderate hydraulic pressure up to 2 GPa exposes contrasting magnetic stability in two Fe-based ladder compounds
with identical crystal structures and similar spin structures. In CsFe$_2$Se$_3$ the stripe-type magnetic phase with $c$-direction spins 
remains unfazed up to the highest measured pressure, while the $a$-direction stripe order in BaFe$_2$S$_3$ goes through a QPT
at about P=1GPa where both the N$\acute{e}$el temperature and the ordered moment abruptly increased. This QPT has the signature 
of an OSMT. Such a finding in a quasi 1D system can narrow down the 
theoretical scope in determining the universal physics that drives the diverse magnetism in iron-based compounds.

Research at Oak Ridge National Laboratory's HFIR was sponsored by the Scientific User Facilities Division, Office of Basic Energy Sciences, U. S. Department of Energy. 
This work was supported by JSPS KAKENHI Grant Number 16H04019. K.O. acknowledges the fruitful discussions with Hiroki Takahashi,  Touru Yamauchi, and Fei Du. 


{}

\end{document}